\documentclass[fleqn,usenatbib]{mnras}
\usepackage[T1]{fontenc}\usepackage{graphicx}\usepackage{amsmath,amssymb,bm}\usepackage{booktabs}\usepackage{newtxtext,newtxmath}
\title[Vortex pinning II]{Vortex pinning and the elastic response of neutron-star crusts -- II. Non-axisymmetric loading and Magnus mountains}
\author[Giliberti]{E. Giliberti$^{1}$\thanks{E-mail: eliagiliberti@gmail.com}\\
$^{1}$Istituto Leonardo da Vinci, Cologno Monzese, Italy}
\date{Submitted to MNRAS -- 1 September 2026}\pubyear{2026}
\begin{document}\pagerange{\pageref{firstpage}--\pageref{lastpage}}\maketitle\label{firstpage}
\begin{abstract}
Pinned superfluid vortices transmit a Magnus force to the neutron-star crust. A non-axisymmetric component of the superfluid--lattice lag can therefore support a persistent mass quadrupole. We calculate the $l=m=2$ response of self-gravitating, radially stratified spherical stars with SLy4 and BSk21 backgrounds, using a local microscopic pinning cap and a density-dependent Coulomb-lattice shear modulus. For a $1.4\,M_\odot$ star at $100\,$Hz, the equilibrium-compression maxima are $\epsilon=2.67\times10^{-9}$ (SLy4) and $2.44\times10^{-9}$ (BSk21). The global quadrupole is strikingly insensitive to the shear prescription: replacing $\mu=0.01P$ by the Coulomb profile and strongly varying uncertain edge layers changes the result only at the percent level. The dominant internal sensitivity is instead compressional. A density-resolved compressional kernel increases through the inner crust and steepens close to the crust--core interface, particularly for BSk21, showing that the deep inner crust controls the EoS dependence of the mountain. A fixed-composition calculation is used only as a non-relaxed sensitivity diagnostic, not as the secular prediction. At a common true local velocity benchmark, our fiducial ellipticities remain more than an order of magnitude below the complementary two-component cylindrical calculation of \citet{GangwarJones2026}. We argue that the comparison points to the relative superfluid--lattice degree of freedom as the natural next ingredient for a common spherical model. The main result is therefore physical rather than numerical: the Magnus-mountain scale is robust to shear microphysics, while the deep-crust compressional response sets the leading uncertainty of the one-displacement model.
\end{abstract}

\section{Introduction}
A steadily rotating neutron star emits continuous gravitational waves if it carries a persistent non-axisymmetric mass quadrupole. Elastic deformations of the crust have therefore long been discussed as possible continuous-wave sources \citep{Ushomirsky2000,Haskell2006,JohnsonMcDaniel2013,Gittins2021,GittinsRel2021}. Vortex pinning provides a distinct route to such a deformation: quantised vortices in the neutron superfluid can remain out of rotational equilibrium with the solid lattice and transmit a Magnus force to it \citep{AnderssonSideryComer2006,HaskellMelatos2015,Seveso2016}. If the lag or pinning pattern contains an $m=2$ component, the reaction force can support a mass quadrupole even when the underlying star is spherical. The stratified elastic--gravitational framework developed by \citet{Giliberti2020} provides the direct methodological precursor to this programme. In Paper~I \citep{PaperI} we applied the same line of reasoning to the axisymmetric mechanical loading generated by the pinned-vortex reaction force; here we ask how much of that loading becomes a non-axisymmetric mountain and which part of the crustal response actually sets its size.

The key distinction is between shear and compression. The shear modulus controls how the solid accommodates shape-changing deformation, whereas the compressional response controls how the forced displacement redistributes mass and hence sources $Q_{22}$. In the inner crust this distinction is entangled with composition and the presence of dripped superfluid neutrons \citep{ChamelHaensel2008,AnderssonSideryComer2006}. We therefore use realistic Coulomb-lattice shear and equilibrium compression for the secular baseline, and use fixed-composition compression only as a controlled non-relaxed sensitivity experiment. This organisation lets the calculation answer a physical question rather than merely compare material prescriptions: which microphysics can genuinely change the Magnus-mountain scale?

A complementary calculation by \citet{GangwarJones2026} retains an explicit neutron--lattice relative degree of freedom in a compressible cylindrical model and finds larger deformations. Our spherical calculation instead resolves self-gravity, realistic radial stratification and crust microphysics with a single displacement field. The difference between the two approaches is therefore useful as a guide to the missing physics, provided the forcing amplitudes are compared consistently. The paper is organised as follows. In Section~2 we formulate the spherical elastic--gravitational response, the pinning-limited Magnus force and the quadrupole normalization. In Section~3 we present the fiducial SLy4 and BSk21 mountains and their continuous-wave scale. Section~4 then asks what controls those mountains: first by separating shear from compressional sensitivity and localising the latter in the deep crust, and then by relating the resulting one-displacement baseline to complementary two-component calculations. We summarise the physical conclusions in Section~5; derivations, normalization details and numerical sensitivity tests are placed in the Appendices.

\section{Physical model}
\subsection{Stellar response}
The spherical Newtonian background satisfies
\begin{align}m'&=4\pi r^2\rho, & P'&=-\rho Gm/r^2.\end{align}
The crust occupies $r_c<r<r_t$. The SLy4 and BSk21 backgrounds used here are constructed from unified crustal descriptions rooted in the corresponding microscopic EoSs \citep{DouchinHaensel2001,Pearson2012} and represented numerically by single analytic pressure--density fits and are hydrostatically integrated as continuous profiles. The radius $r_c$ is selected by the adopted transition density; it is not implemented as a two-density discontinuity. Consequently the numerical model has $[\rho]_{r_c}=0$ and contains no delta-function mass sheet at the fluid--solid interface. The only material discontinuity imposed at $r_c$ is the onset of the shear modulus. This distinction is important for the gravitational matching described in Appendix~\ref{app:system}. For each spheroidal harmonic,
\begin{equation}\bm\xi_{lm}=U_{lm}(r)Y_{lm}\hat{\bm r}+V_{lm}(r)\nabla_\perp Y_{lm}.\end{equation}
We integrate the six-component state
\begin{equation}\bm y=(U,V,R,S,\Phi,Q),\qquad Q=\Phi'+\frac{l+1}{r}\Phi+4\pi G\rho U.\label{eq:state}\end{equation}
with
\begin{equation}\bm y'=\bm A(r,l)\bm y-\bm h(r).\label{eq:firstorder}\end{equation}
Here $R$ and $S$ are radial and tangential traction amplitudes. The explicit matrix, force projection and boundary conditions are given in Appendix~\ref{app:system}; this removes any ambiguity in the reproducibility of the calculation.

\subsection{Elastic and compressional microphysics}\label{sec:mu}
The production calculation uses the standard zero-temperature orientationally averaged Coulomb-lattice shear modulus \citep{OgataIchimaru1990,Strohmayer1991,ChamelHaensel2008}
\begin{equation}
 \mu_{\rm C}(r)=0.1194\,n_i(r)\frac{[Z(r)e]^2}{a(r)},\qquad
 a=\left(\frac{3}{4\pi n_i}\right)^{1/3}.
 \label{eq:muc}
\end{equation}
The ion density and charge are taken from the adopted crust composition. For BSk21 the baseline uses the Wigner--Seitz-cell proton number $Z'=40$, as found throughout the inner crust in the TETFSI BSk21 calculation of \citet{Pearson2012}; using the localized cluster charge instead as an alternative deep-crust prescription changes the maximum ellipticity by only about one per cent. Equation~(\ref{eq:muc}) is inserted directly into the elastic operator through $\lambda=K-2\mu/3$ and $\beta=K+4\mu/3$: none of the quadrupoles quoted below is obtained by rescaling a solution computed with $\mu=0.01P$.

The distinction is also separate from the breaking-strain problem: the crust may be very strong locally \citep{HorowitzKadau2009}, while the global quadrupole supported by a specified force can remain small. This distinction matters locally. In much of the inner crust $\mu_{\rm C}$ is only about one half of $0.01P$, so the two prescriptions represent materially different elastic profiles. Nevertheless the global $l=m=2$ response is remarkably insensitive to this change. We exploit that fact below as a physical diagnostic rather than as a numerical convenience.

\subsubsection{Equilibrium and fixed-composition compression}\label{sec:frozen}
The density perturbation produced by a given Magnus load depends on the compressional closure. In chemical equilibrium we use
\begin{equation}
 K_{\rm eq}=n_{\rm B}\frac{dP_{\rm eq}}{dn_{\rm B}}.
\end{equation}
For loading short compared with the relevant composition-relaxation times, the local particle fractions cannot readjust and the appropriate limiting derivative is instead taken at fixed composition. We denote the corresponding modulus by $K_f$ and impose $K_f\ge K_{\rm eq}$ in the reconstructed profile. The secular Magnus mountain considered here is expected to build on the lag-accumulation timescale, so $K_{\rm eq}$ is our primary baseline; $K_f$ is used as a limiting non-relaxed response, relevant to assessing the sensitivity to faster loading episodes rather than as a second co-equal secular prediction.

For the present calculation we use EoS-specific fixed-composition profiles reconstructed point by point for SLy4 and BSk21 over
\begin{equation}
 0.01\le n_{\rm B}\le0.06\ {\rm fm^{-3}}.
 \label{eq:matched-domain}
\end{equation}
This is the interval over which the fixed-composition reconstruction is microphysically controlled. Outside it we return to $K_{\rm eq}$; in particular, we do not extrapolate the frozen correction through the remaining layer to $n_{cc}=0.0800\,{\rm fm^{-3}}$ (SLy4) or $0.0809\,{\rm fm^{-3}}$ (BSk21). The transition at each edge of Eq.~(\ref{eq:matched-domain}) is smoothed over $0.002\,{\rm fm^{-3}}$. This width is a numerical regularisation, not a microphysical transition scale. Replacing the PCHIP representation by linear interpolation, or varying/removing the taper, changes the resulting ellipticity by sub-percent amounts. We therefore refer to this calculation as the \emph{matched-domain frozen} limiting response. Because the density-resolved quadrupolar susceptibility increases sharply beyond $0.06\,\mathrm{fm^{-3}}$ (Section~\ref{sec:qkernel}), the unmodelled continuation to the crust--core boundary is a genuine component of the theoretical uncertainty of the non-relaxed response. We do not assign it a sign or convert the matched-domain result into a bound, since the fixed-composition profile in that layer has not been calculated. The equilibrium-composition solution remains our fiducial prediction for a secularly accumulated pinning mountain.

This construction is deliberately conservative. It asks how much of the Magnus mountain changes when the part of the crust for which the fixed-composition thermodynamics is known is allowed to respond in that limit, while leaving the unresolved deepest layer untouched. The density-resolved kernel in Section~\ref{sec:qkernel} is then used to show why this distinction matters physically.

\subsection{Pinning-limited Magnus forcing}
The vortex areal density is $n_v=2\Omega/\kappa$, with circulation quantum $\kappa=h/(2m_n)\simeq1.99\times10^{-3}\,{\rm cm^2\,s^{-1}}$. For lag $\Delta\Omega$ the characteristic Magnus force per unit vortex length is
\begin{equation}f_{\rm M,line}=\rho_s\kappa r\sin\theta\,\Delta\Omega.\end{equation}
We impose the microscopic pinning limit locally,
\begin{equation}f_{\rm line}(r,\theta,\phi)=\min[\rho_s\kappa r\sin\theta\,\Delta\Omega(\phi),f_{\rm pin}(r)],\end{equation}
and $f_{\rm vol}=n_v f_{\rm line}$. The radial pinning profile follows the mesoscopic scale used in our numerical implementation, based on \citet{Seveso2016}. We adopt
\begin{equation}\Delta\Omega(\phi)=\Delta\Omega_0[1+\eta_2\cos(2\phi)],\quad 0\leq\eta_2\leq1,\end{equation}
with $\eta_2=1$ as an upper-envelope $m=2$ asymmetry. The cap is applied before Fourier and spherical-harmonic projection; therefore the $m=2$ force ceases to scale linearly once parts of the crust saturate.

\subsection{Quadrupole normalization}
We use the real, unnormalised angular basis
\begin{equation}\delta\Phi=\Phi^u_{22}(r)P_2^2(\cos\theta)\cos2\phi,\qquad P_2^2=3\sin^2\theta.\end{equation}
This is the associated Legendre function in the Condon--Shortley convention for $l=m=2$; no unit-normalised $Y_{22}$ factor is included in $\Phi^u_{22}$. With this convention,
\begin{align}I_{xx}-I_{yy}&=\frac{4r_t^3}{G}\Phi^u_{22}(r_t),\\
\epsilon&=\frac{4r_t^3}{GI_{zz}}|\Phi^u_{22}(r_t)|.
\end{align}
The same quadrupole is reconstructed independently from the Eulerian density perturbation, including core and moving-boundary terms; an independent rebuild reproduces the equilibrium ellipticity at the $\sim10^{-4}$ relative level.

\section{Magnus mountains and continuous-wave scale}
\subsection{Fiducial mountains}
For $M=1.4\,M_\odot$, $f_{\rm rot}=100\,$Hz and the maximal quadrupolar asymmetry $\eta_2=1$, the Coulomb-shear equilibrium-compression solutions reach
\begin{align}
\epsilon_{\rm eq}^{\rm SLy4}&=2.671\times10^{-9}, & \Delta\Omega_{0,\max}&=1.488\times10^{-2}\,{\rm rad\,s^{-1}},\\
\epsilon_{\rm eq}^{\rm BSk21}&=2.438\times10^{-9}, & \Delta\Omega_{0,\max}&=1.418\times10^{-2}\,{\rm rad\,s^{-1}}.
\end{align}
These are the fiducial secular predictions used throughout the observational and Gangwar--Jones comparisons. Their proximity is itself informative: once the same realistic shear prescription and equilibrium compressional closure are used, the two EoSs support very similar global mountains despite their different radial structure.

\begin{table}
\centering
\caption{Fiducial maximum pinning-supported ellipticity at $100\,$Hz for a $1.4\,M_\odot$ star, using realistic Coulomb shear and equilibrium compression.}
\label{tab:main}
\begin{tabular}{lcc}
\toprule
EoS & $\epsilon_{\max}$ & $\Delta\Omega_{0,\max}\,[{\rm rad\,s^{-1}}]$\\
\midrule
SLy4  & $2.671\times10^{-9}$ & $1.488\times10^{-2}$\\
BSk21 & $2.438\times10^{-9}$ & $1.418\times10^{-2}$\\
\bottomrule
\end{tabular}
\end{table}

The values in Table~\ref{tab:main} are the secular baseline. At the common true-local-velocity benchmark $\max|\delta v|=10^4\,{\rm cm\,s^{-1}}$, they become $1.55\times10^{-9}$ (SLy4) and $1.37\times10^{-9}$ (BSk21), which we use only when placing different model conventions on the same local-velocity scale in Section~\ref{sec:GJ}. The fixed-composition response and numerical convergence checks are discussed later as diagnostics of the physical and numerical uncertainty, not as additional predictions.

\subsection{Continuous-wave scale and high-spin limit}
For principal-axis rotation, using the standard continuous-wave amplitude convention \citep{Prix2004},
\begin{equation}h_0=\frac{16\pi^2G}{c^4}\frac{I_{zz}\epsilon f_{\rm rot}^2}{d}.\end{equation}
Within a fixed spherical background the vortex density scales as $\Omega$, and the maximum pinning-supported ellipticity scales approximately linearly with spin frequency. The O4a targeted search reports a lowest ellipticity upper limit $\epsilon^{95\%}=8.8\times10^{-9}$ for J0437$-$4715 \citep{LVKO4a2025}. The more recent O4a+O4b narrowband analysis reaches $1.1\times10^{-8}$ for J0711$-$6830 and $1.7\times10^{-8}$ for the first J1400$-$1431 segment \citep{LVKO4ab2026}.

\begin{figure}\centering\includegraphics[width=\columnwidth]{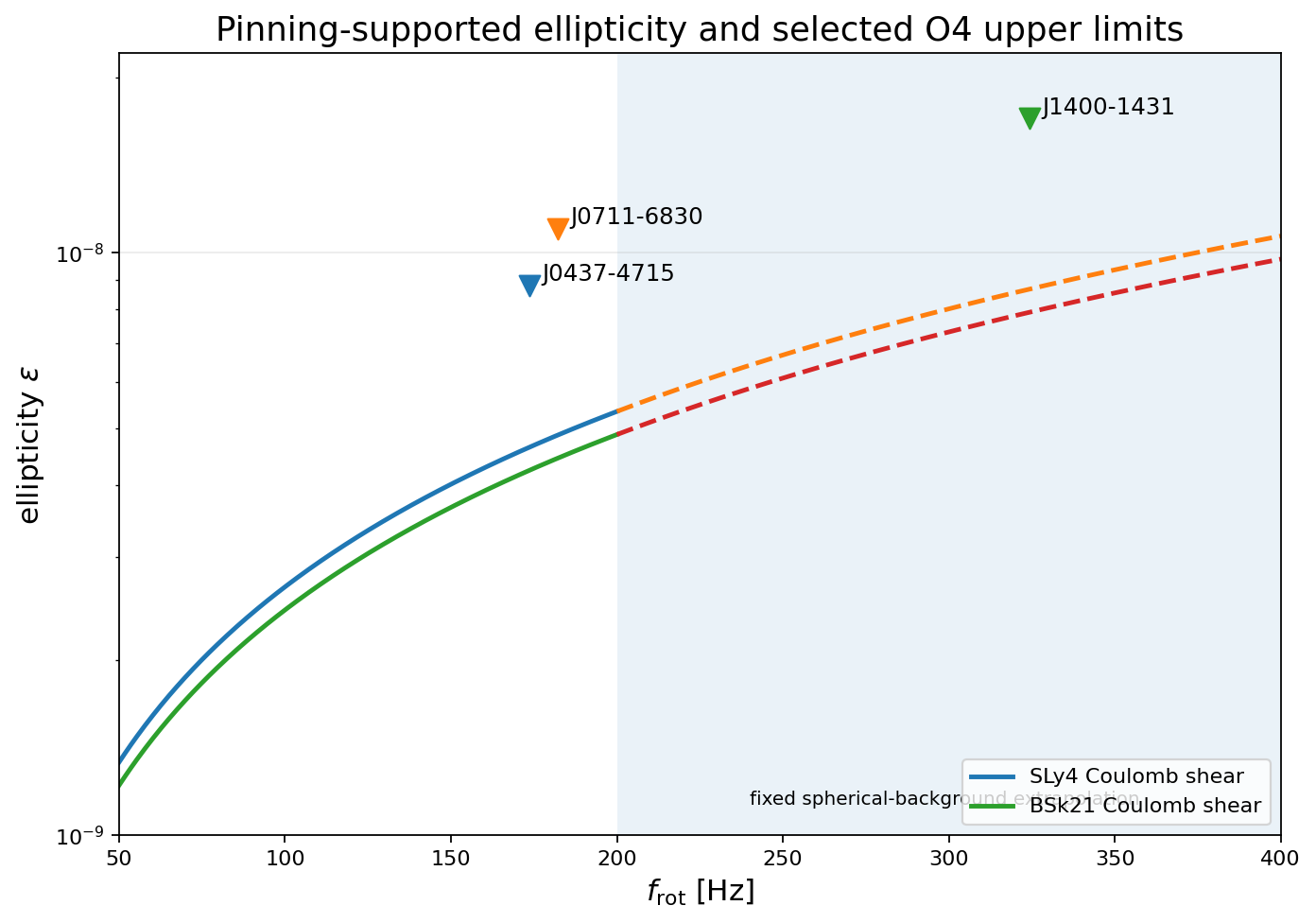}\caption{Maximum pinning-supported ellipticity versus spin frequency compared with selected O4 upper limits. Pulsar markers are placed at their observed spin frequencies. The theory curves are solid up to $200$ Hz and dashed above it; the shaded region marks the same fixed-spherical-background extrapolation. Rotational flattening, background--perturbation mode coupling and relativistic corrections are not included, so this part of the curve is an observational-reach diagnostic rather than a precision prediction.}\label{fig:cw}\end{figure}

Figure~\ref{fig:cw} is an observational-reach diagnostic, not a rapidly rotating stellar sequence. We conservatively flag the plotted theory curves above 200 Hz as fixed-spherical-background extrapolations. In this regime the equilibrium centrifugal deformation is no longer a parametrically negligible correction to the background geometry, and couplings among the rotationally deformed background and the forced $l=2$ response are omitted by construction. The large axisymmetric rotational quadrupole does not by itself imply a comparably large spurious $m=2$ mountain, but the cross-coupling has not been calculated here; therefore the high-spin curves must not be read as precision predictions for individual millisecond pulsars. Centrifugal flattening, rotational mode coupling and relativistic structure are omitted. There is no model-independent expectation for the sign of the oblateness correction in this problem. An oblate background changes the local crustal thickness, density surfaces, vortex geometry and pinning-weighted lever arm differently between equator and pole, so competing effects can either enhance or reduce the $m=2$ response. Determining the sign requires solving the Magnus-forced perturbation on a self-consistently rotating background rather than extrapolating the spherical sequence. We do not assign a universal percentage error to these omissions because it has not been calculated for the present Magnus-forced boundary-value problem.

\section{What controls the Magnus mountain?}
\subsection{From shear robustness to compressional leverage}
The weak dependence on $\mu(r)$ is physically informative. The Coulomb-lattice profile can differ substantially from the commonly used proxy $0.01P$ throughout the inner crust, yet the equilibrium maximum changes by only a few per cent. Even deliberately large changes confined to the outer crust or to a putative pasta layer alter $Q_{22}$ at the percent level or below. The reason is that the Magnus mountain is not a local strain estimate: changing $\mu(r)$ redistributes the displacement and density perturbation through the coupled elastic--gravitational problem, while the integrated $m=2$ mass moment is comparatively stable. The numerical variants supporting this statement are collected in Appendix~\ref{app:shearchecks}. This robustness already shows that shear microphysics alone cannot account for the larger deformation found in the two-component calculation discussed next.

This weak shear dependence shifts the question from local rigidity to mass redistribution. The fiducial calculations expose a useful separation of roles. Shear determines how the solid accommodates the Magnus load locally, but plausible changes in $\mu(r)$ barely move the integrated quadrupole. Compression acts more directly on the density perturbation that sources $Q_{22}$. A stiffer local compressional response does not simply reduce the deformation: it reorganises $U$, $V$, $\delta\rho$ and $\delta\Phi$ throughout the star, so the global response depends on where in the crust the change occurs. This motivates a density-resolved measure of compressional leverage rather than a single effective adiabatic index.

\subsection{The deep crust as the mechanical lever arm}\label{sec:qkernel}
\begin{figure}
\centering
\includegraphics[width=\columnwidth]{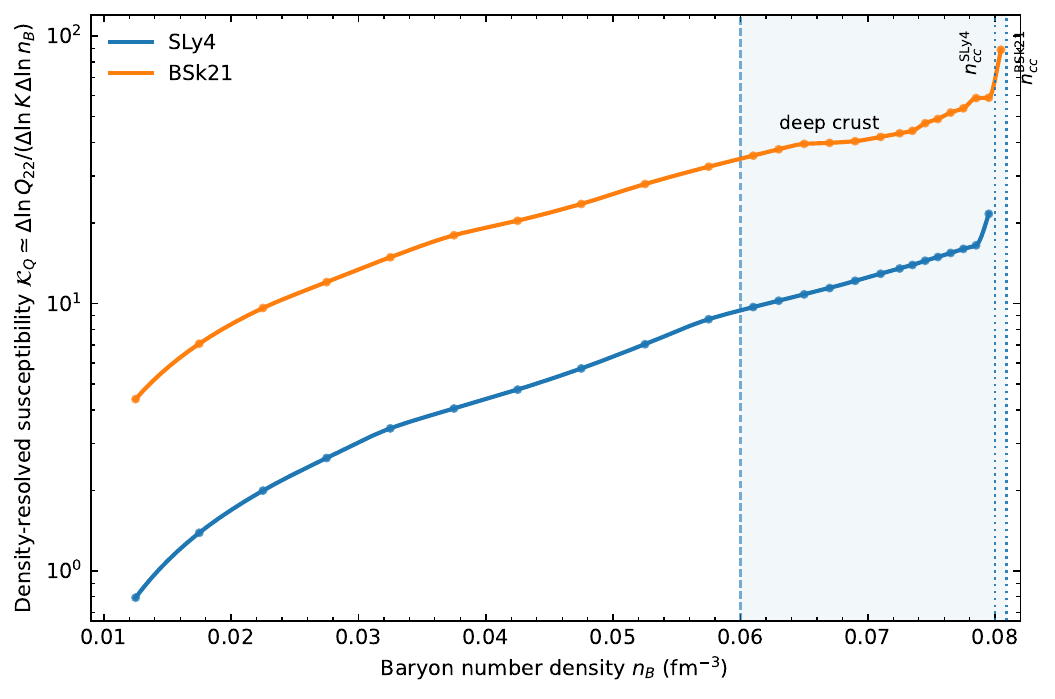}
\caption{Density-resolved compressional susceptibility ${\cal K}_Q$ [Eq.~(\ref{eq:kernel_density})] for SLy4 and BSk21. Points are independent symmetric $\pm0.5$ per cent perturbations of $K$ in narrow density windows; solid curves are shape-preserving interpolants through the numerical samples. The shaded region begins at $n_B=0.06\,{\rm fm^{-3}}$, beyond the microphysically controlled matched-domain frozen profile. Dotted vertical lines mark the EoS-specific crust--core interfaces. The response grows through the inner crust and steepens close to the interface, with a substantially larger susceptibility for BSk21.}
\label{fig:qkernel}
\end{figure}

The weak shear dependence points to a more useful question: \emph{where} does a change in compression most efficiently alter the mass quadrupole? We answer it by perturbing $K$ in progressively narrower density windows while keeping the Magnus forcing, shear modulus and remaining material properties fixed. For a finite window $[n_1,n_2]$ we first compute the usual logarithmic response $S_Q=\Delta\ln Q_{22}/\Delta\ln K$. To compare windows of unequal width without turning the last, wider interval into an artificial ``peak'', Fig.~\ref{fig:qkernel} shows the response per logarithmic density interval,
\begin{equation}
 {\cal K}_Q(n_B)\simeq
 \frac{\Delta\ln Q_{22}}
 {\Delta\ln K\,\Delta\ln n_B},
 \qquad \Delta\ln n_B=\ln(n_2/n_1).
 \label{eq:kernel_density}
\end{equation}
Each point is obtained from a symmetric $\pm0.5$ per cent perturbation of $K$. The sampling is refined from $\Delta n_B=0.005\,{\rm fm^{-3}}$ in the shallower inner crust to $0.002\,{\rm fm^{-3}}$ above $0.06\,{\rm fm^{-3}}$ and $0.001\,{\rm fm^{-3}}$ close to the crust--core interface. The lines in Fig.~\ref{fig:qkernel} are shape-preserving interpolants through these calculated points, not additional model fits.

The higher-resolution calculation changes the interpretation of the earlier coarse-bin view in a useful way. The deep-crust enhancement is real, but it is not an isolated spike generated at the centre of the terminal bin. The susceptibility increases progressively through the inner crust for both EoSs and steepens again immediately before the crust--core interface. BSk21 remains systematically more responsive than SLy4 throughout the sampled region. Thus the physically robust statement is not that one particular bin dominates, but that the compressional leverage is increasingly concentrated toward the base of the crust. The two EoS-specific interfaces, $n_{cc}=0.0800\,{\rm fm^{-3}}$ for SLy4 and $0.0809\,{\rm fm^{-3}}$ for BSk21, are marked explicitly in Fig.~\ref{fig:qkernel}.

This localisation clarifies the role of the fixed-composition experiment. Applied only over the controlled interval $0.01\le n_B\le0.06\,{\rm fm^{-3}}$, it raises the solved ellipticity by $7.5$ per cent for SLy4 and $20.6$ per cent for BSk21. These are not alternative secular predictions. Rather, they demonstrate that changing the compressional response in a region to which the quadrupole is already sensitive produces a measurable global shift, while the still more responsive layer approaching the interface is deliberately left at equilibrium compression. This is why extrapolating the frozen correction to $n_{cc}$ would be unjustified even though it could have appreciable leverage.

The stronger BSk21 response should not be assigned to a single bulk nuclear parameter. SLy4 and BSk21 differ simultaneously in inner-crust composition, equilibrium compressibility, pressure gradient, crust thickness and the mapping between density and radius; these enter the coupled elastic--gravitational response together. The present comparison therefore supports a narrower physical conclusion: the EoS dependence of the one-displacement Magnus mountain is controlled primarily by how deep-crust compression couples to global mass redistribution. Isolating a causal dependence on, for example, the symmetry-energy slope would require a controlled EoS family rather than two isolated models.

This result also provides the natural bridge to the next question. If realistic changes in shear leave the mountain almost unchanged, while compression can move it appreciably but still within the same one-displacement framework, can the much larger quadrupoles found in models with an explicit neutron--lattice relative degree of freedom be explained by material microphysics alone? We address that comparison next.

\subsection{Relation to complementary two-component models}\label{sec:GJ}
\citet{GangwarJones2026} describe the same basic mechanism with a complementary set of approximations: their model retains an explicit neutron--lattice relative degree of freedom in a cylindrical geometry, whereas ours resolves spherical stratification, self-gravity and crust microphysics with a single displacement field. A direct comparison is therefore most useful when the forcing amplitudes are placed on the same footing.

Their quoted velocity parameter is not the maximum local magnitude of the prescribed relative-velocity field. Using their fiducial cylindrical annulus, the conversion from their amplitude convention to the true local maximum gives a shape factor $C_{\rm sh}=153.5$. Thus imposing the same local benchmark used here, $\max|\delta\bm v|=10^4\,{\rm cm\,s^{-1}}$, maps their 100-Hz fit to $\epsilon_{\rm GJ}^{\rm local}=5.60\times10^{-8}$. Our equilibrium Coulomb-shear solutions at that benchmark are $1.55\times10^{-9}$ (SLy4) and $1.37\times10^{-9}$ (BSk21), leaving factors of about 36 and 41, respectively.

This comparison is most useful as a scale comparison rather than as a single exact ratio. The conversion depends on the adopted cylindrical geometry, and varying the annulus around the fiducial choice changes it substantially; throughout the range tested, however, the two calculations remain separated by more than an order of magnitude. The derivation and boundary scan are given in Appendix~\ref{app:GJnorm}. A Jones-like spherical toy model also remains well above the realistic SLy4/BSk21 solutions, so replacing cylindrical geometry by spherical geometry alone does not close the gap.

We therefore use the comparison diagnostically. Realistic shear does not remove the separation, and the compressional tests below move the one-displacement solution without bringing it to the two-component scale. The most consequential structural difference is then the degree of freedom that our model does not yet contain: independent motion of the superfluid neutrons relative to the lattice, together with the associated thermodynamic closure and entrainment. In this sense the two calculations are complementary limits of a more complete problem, rather than competing estimates that should already agree numerically.

\begin{figure}
\centering
\includegraphics[width=\columnwidth]{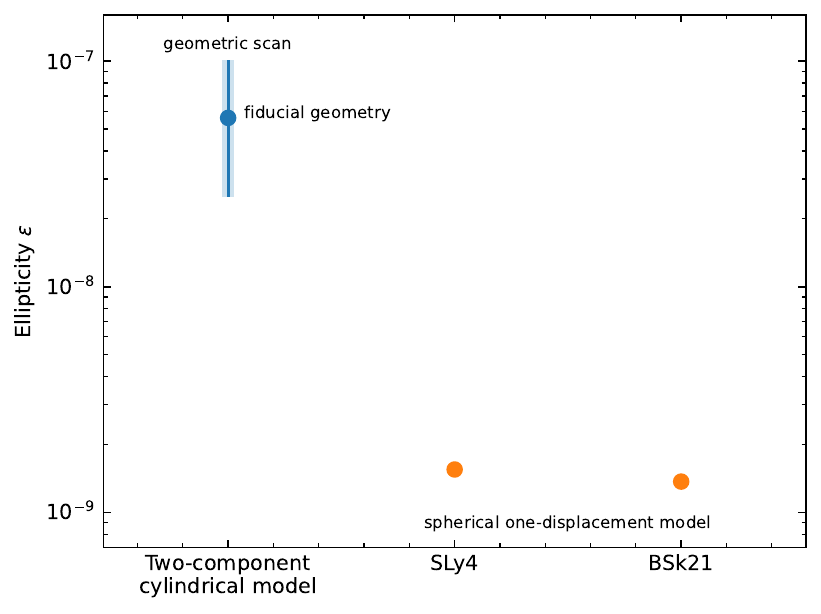}
\caption{Ellipticity at $100$ Hz on a common true-local-velocity benchmark. The left interval shows the two-component cylindrical model of \citet{GangwarJones2026}: the point is their fiducial geometry after conversion to the common local cap, while the vertical interval shows the sensitivity of that cross-model normalization to the small annulus-boundary scan described in Appendix~\ref{app:GJnorm}. The SLy4 and BSk21 points are the present spherical one-displacement predictions. The interval is a normalization-sensitivity diagnostic, not a statistical uncertainty. Even across the scan, the two-component result remains more than an order of magnitude above the present solutions.}
\label{fig:GJ}
\end{figure}

Seen in this order, the comparison does not introduce a second, disconnected result. It asks whether the microphysics identified above can bridge the remaining difference between a realistic spherical one-displacement star and a model with explicit relative superfluid motion. The answer is no: shear changes too little, while controlled compressional changes are significant but do not reach the two-component scale. This makes the missing relative degree of freedom the natural physical target of the next calculation.

The same comparison also points naturally beyond a one-displacement description. In the inner crust, dripped neutrons and the charged lattice need not share the same displacement, and entrainment alters the relation between transport velocities and momenta \citep{AnderssonSideryComer2006,ChamelHaensel2008}. A reduced static thermodynamic closure can be written through the Schur complement $D=E_{pp}-E_{np}^2/E_{nn}$, with $K_{\rm mf}=n_c^2D$, but this thermodynamic reduction cannot replace the missing relative displacement. The decisive next calculation is therefore spherical and stratified like the present model while retaining independent neutron and lattice degrees of freedom. In this sense, the present solutions provide the elastic--gravitational baseline against which a common multifluid extension can be tested.

\subsection{Domain of validity}
The quantitative results are Newtonian and based on spherical backgrounds. The $1.4\,M_\odot$ model is deliberately used as the fiducial mass throughout Paper~II; a full mass sequence of non-axisymmetric quadrupoles is deferred because changing $M$ simultaneously changes the radius, crust thickness, density--radius mapping and pinning-weighted local-cap geometry. The spin-frequency curves in Fig.~\ref{fig:cw} are therefore best regarded as observational-reach diagnostics. Above $200\,$Hz we explicitly flag the fixed-background extrapolation because rotational flattening, mode coupling on an oblate background and relativistic structure are omitted. We do not assign a sign to these corrections without solving the forced problem on a rotating background.

The matched-domain frozen calculation has a separate, clearly defined limitation: it is a controlled fixed-composition limiting response only over Eq.~(\ref{eq:matched-domain}). Since the quadrupolar kernel becomes largest closer to the crust--core boundary, a full frequency- and temperature-dependent relaxation calculation through the complete inner crust is the physically relevant extension. The equilibrium values in Table~\ref{tab:main} are our secular baseline; the frozen values quantify a non-relaxed sensitivity and are not presented as co-equal secular predictions or as bounds on the unknown kinetic response.

\section{Conclusions}
We have calculated the non-axisymmetric Magnus mountain produced by pinned vortices in spherical, self-gravitating SLy4 and BSk21 stars. For a $1.4\,M_\odot$ star at $100\,$Hz, realistic Coulomb shear and equilibrium compression give $\epsilon_{\max}=2.67\times10^{-9}$ and $2.44\times10^{-9}$, respectively. These are our secular predictions. Their most striking property is not their absolute size but their robustness: substantial changes in the local shear prescription alter the global quadrupole only weakly.

The dominant internal leverage lies instead in the compressional response of the deep crust. The refined density-resolved kernel grows progressively through the inner crust and steepens close to the crust--core interface, while remaining substantially larger for BSk21. A matched fixed-composition experiment confirms that chemically unrelaxed compression can move the solution appreciably, but it is retained only as a sensitivity diagnostic. The physical picture is therefore that a Magnus mountain is set by the radial coupling between the forcing, compressional microphysics and self-gravity, rather than by a single measure of crustal rigidity.

At a common local-velocity benchmark, the fiducial Gangwar--Jones two-component result remains more than an order of magnitude above the present one-displacement solutions. We do not regard this as a conflict: the two calculations resolve different ingredients. The comparison identifies the natural next step -- a spherical, stratified model with independent superfluid and lattice displacements and consistent entrainment. Paper~II thus establishes the realistic one-displacement baseline and, more importantly, isolates the deep-crust compressional physics that controls where further microphysics can change the answer.

\appendix
\section{Explicit radial operator and boundary conditions}\label{app:system}
Define $L=l(l+1)$, $\lambda=K-2\mu/3$ and $\beta=K+4\mu/3$. In the dimensionless implementation used for the numerical integration (with $G=1$ after stellar scaling), the non-zero entries of the matrix in Eq.~(\ref{eq:firstorder}) are
\begin{align}
A_{11}&=-\frac{2\lambda}{r\beta},&A_{12}&=\frac{L\lambda}{r\beta},&A_{13}&=\beta^{-1},\\
A_{21}&=-r^{-1},&A_{22}&=r^{-1},&A_{24}&=\mu^{-1},\\
A_{31}&=\frac{4}{r}\left(\frac{3K\mu}{r\beta}-\rho g\right),&
A_{32}&=\frac{L}{r}\left(\rho g-\frac{6K\mu}{r\beta}\right),\\
A_{33}&=-\frac{4\mu}{r\beta},&A_{34}&=\frac{L}{r},\\
A_{35}&=-\frac{\rho(l+1)}{r},&A_{36}&=\rho,\\
A_{41}&=\frac{1}{r}\left(\rho g-\frac{6\mu K}{r\beta}\right),&
A_{42}&=\frac{2\mu}{r^2}\left[L\left(1+\frac{\lambda}{\beta}\right)-1\right],\\
A_{43}&=-\frac{\lambda}{r\beta},&A_{44}&=-\frac{3}{r},&A_{45}&=\frac{\rho}{r},\\
A_{51}&=-4\pi\rho,&A_{55}&=-\frac{l+1}{r},&A_{56}&=1,\\
A_{61}&=-\frac{4\pi\rho(l+1)}{r},&A_{62}&=\frac{4\pi\rho L}{r},&A_{66}&=\frac{l-1}{r}.
\end{align}
All omitted entries vanish. Restoring dimensions amounts to applying the stellar length, density, pressure and force-density scales used in the background nondimensionalisation. The entries above have been checked term by term against the first-order variables defined in Eq.~(\ref{eq:state}); throughout this appendix $l$ is the spherical-harmonic degree, $L\equiv l(l+1)$, and matrix subscripts refer only to the ordering $(U,V,R,S,\Phi,Q)$.

For a spheroidal body force
\begin{equation}\bm f=f_r(r,\theta,\phi)\hat{\bm r}+f_\perp(r,\theta,\phi)\hat{\bm e}_\perp,\end{equation}
the radial and tangential forcing coefficients are obtained by angular projection before radial integration. In the axisymmetric Legendre implementation used to validate the operator,
\begin{align}h_R^{(l)}&=\frac{2l+1}{2}\int_{-1}^{1}f_rP_l(x)\,dx,\\
h_S^{(l)}&=\frac{2l+1}{2l(l+1)}\int_{-1}^{1}f_\theta\,\partial_\theta P_l\,dx,\end{align}
with $h_S^{(0)}=0$; the non-axisymmetric calculation uses the corresponding real $m=2$ spherical-harmonic projection. With the sign convention of Eq.~(\ref{eq:firstorder}), $\bm h=(0,0,h_R,h_S,0,0)$.

At the fluid--solid interface the fluid carries no shear traction, hence
\begin{equation}S(r_c)=0.\end{equation}
Radial traction and gravitational potential are matched to the regular fluid-core solution. In the production backgrounds used in this paper, $\rho(r)$ is continuous at $r_c$ because SLy4 and BSk21 are represented by unified analytic pressure--density fits and $r_c$ is selected as an exact transition-density node. Hence $[\rho]_{r_c}=0$, no explicit surface-mass term proportional to $[\rho]U$ is present, and the limiting values of $\Phi'$ are continuous. The variable $Q=\Phi'+(l+1)\Phi/r+4\pi G\rho U$ is therefore matched without an additional density-jump term. This should not be confused with a model containing a first-order density discontinuity: for such a background the corresponding jump conditions for $\Phi'$ and radial traction would have to be added explicitly. The present calculation does not make that approximation. The shear modulus does turn on at the fluid--solid interface; an exact crust-side node is used so that the elastic operator never interpolates across $\mu=0$ in the first crustal step. At the adopted elastic top, the solid is traction free,
\begin{equation}R(r_t)=0,\qquad S(r_t)=0,\end{equation}
and the gravitational perturbation is matched to the decaying exterior solution $\Phi_{\rm ext}\propto r^{-(l+1)}$. Therefore
\begin{equation}Q(r_t)=0\end{equation}
when no external forcing potential is present. The numerical solution is built from regular core basis solutions plus a forced particular solution and solves the three outer conditions $(R,S,Q)=(0,0,0)$ simultaneously. These are the boundary conditions used for the pinning calculation.

\section{Crust--core interface audit}
The interface prescription was checked against the background builder used by the production solver. The hydrostatic profile is obtained from a single-valued $P(\rho)$ relation on both sides of $r_c$; $r_c$ is then located by the transition-density criterion ($n_{cc}=0.0800\,\mathrm{fm^{-3}}$ for SLy4 and $0.0809\,\mathrm{fm^{-3}}$ for BSk21 in the adopted builder). Thus the left and right density limits coincide by construction. For the reconstructed $1.4\,M_\odot$ backgrounds we obtain $R=13.7241$ km and $r_c/R=0.86436$ for SLy4, and $R=14.2672$ km and $r_c/R=0.86851$ for BSk21. There is therefore no adjustable density-smoothing width whose variation could generate a hidden $Q_{22}$ systematic in the present model. A separate physical question is the sharp onset of $\mu$; earlier solver robustness tests used a smoothed shear transition as a numerical diagnostic, while the production solution uses an exact crust-side interface node. Because the current paper does not model a first-order phase-transition density jump, no claim is made about the sensitivity of $Q_{22}$ to such a jump.

\section{Shear-profile sensitivity checks}\label{app:shearchecks}
For completeness, Table~\ref{tab:musysapp} collects the material variations used to verify that the global quadrupole is not controlled by uncertain shear physics at the edges of the elastic crust. The entries are changes relative to the Coulomb-shear baseline.
\begin{table}
\centering
\caption{Sensitivity of the 100-Hz maximum quadrupole to the shear-modulus prescription.}
\label{tab:musysapp}
\begin{tabular}{lcc}
\toprule
Modification & SLy4 [\%] & BSk21 [\%]\\
\midrule
$\mu=0.01P$ & $-4.2$ & $+0.8$\\
outer crust $\times0.5$ & $+0.01$ & $-0.03$\\
outer crust $\times2$ & $-0.02$ & $+0.04$\\
pasta taper to $0.5\mu_{\rm C}$ & $+0.83$ & $-0.50$\\
pasta taper to $0.1\mu_{\rm C}$ & $+1.65$ & $-1.03$\\
\bottomrule
\end{tabular}
\end{table}
The signs here refer to replacing the Coulomb baseline by the listed variant. Conversely, describing the change from the $0.01P$ proxy to the Coulomb profile reverses the sign, giving $+4.4$ per cent for SLy4 and $-0.8$ per cent for BSk21. The outer-crust and pasta variations are intentionally broad numerical brackets rather than claims for specific microphysical transition profiles.

\section{Gangwar--Jones velocity normalization}\label{app:GJnorm}
The comparison in Section~\ref{sec:GJ} uses the velocity field defined by \citet{GangwarJones2026}. With $x=r/R$, their fiducial annulus has $x_{\rm in}=0.9$ and $x_{\rm out}=0.99$, and the radial functions entering the prescribed flow are
\begin{align}
 u(x)&=2x\left(\frac{x^2}{x_{\rm out}^2}-1\right)\left(\frac{x^2}{x_{\rm in}^2}-1\right),\\
 v(x)&=\frac{6x^5}{x_{\rm out}^2x_{\rm in}^2}-\frac{4x^3}{x_{\rm out}^2}-\frac{4x^3}{x_{\rm in}^2}+2x.
\end{align}
Their fitted amplitude convention and local field imply
\begin{equation}
 C_{\rm sh}=\frac{\max|\delta\bm v_n|}{\delta v_n}=\max\left[\frac{\rho_{n,0}}{0.2\rho_n(x)}\sqrt{u^2+v^2}\right].
\end{equation}
For the fiducial geometry $C_{\rm sh}=153.5$, giving $\epsilon_{\rm GJ}^{\rm local}=5.60\times10^{-8}$ at the common $10^4\,{\rm cm\,s^{-1}}$ local cap. A small boundary scan, $x_{\rm in}=0.89$--0.91 and $x_{\rm out}=0.985$--0.995, gives $85.1\le C_{\rm sh}\le342.2$ and hence $2.51\times10^{-8}\le\epsilon_{\rm GJ}^{\rm local}\le1.01\times10^{-7}$. This scan is a sensitivity test of our cross-model normalization, not an uncertainty distribution for the Gangwar--Jones model.

\section{Numerical sensitivity audit}
An independent rebuild of the corrected $m=2$ calculation reproduces the equilibrium ellipticities to $+0.0170$ per cent (SLy4) and $-0.0125$ per cent (BSk21). The density-resolved stiffness audit is recomputed with the production Coulomb shear and perturbs one density band at a time while holding the forcing and remaining material properties fixed. The quoted $S_Q$ values use a symmetric $\pm0.5$ per cent perturbation in $K$; additional $0.5$, $1$ and $2$ per cent one-sided checks confirm the trend and show increasing nonlinearity in the deepest bands. These tests are designed to measure the response of the solved boundary-value problem and are not interpreted as reconstructed whole-profile microphysical closures.

\section*{Data availability}
Machine-readable tables for the fiducial equilibrium runs, the controlled fixed-composition sensitivity test, the realistic-shear systematics, the Gangwar--Jones normalization factor, and the Coulomb-shear density-resolved compressional kernel are included in the submission data package, together with the Python scripts used for the reruns. The radial operator, forcing convention and interface conditions required to reproduce the boundary-value problem are given in the text and Appendix~A.

\label{lastpage}
\begin{thebibliography}{99}
\bibitem[Abac et al.(2025)]{LVKO4a2025} Abac A. G., et al., 2025, ApJ, 983, 99
\bibitem[Andersson, Sidery \& Comer(2006)]{AnderssonSideryComer2006} Andersson N., Sidery T., Comer G. L., 2006, MNRAS, 368, 162
\bibitem[Chamel \& Haensel(2008)]{ChamelHaensel2008} Chamel N., Haensel P., 2008, Living Rev. Relativ., 11, 10
\bibitem[Douchin \& Haensel(2001)]{DouchinHaensel2001} Douchin F., Haensel P., 2001, A\&A, 380, 151
\bibitem[Gangwar \& Jones(2026)]{GangwarJones2026} Gangwar Y., Jones D. I., 2026, arXiv:2608.12378
\bibitem[Giliberti et al.(2020)]{Giliberti2020} Giliberti E., Cambiotti G., Antonelli M., Pizzochero P. M., 2020, MNRAS, 491, 1064
\bibitem[Giliberti \& Cambiotti(2026)]{PaperI} Giliberti E., Cambiotti G., 2026, arXiv:2608.30029
\bibitem[Gittins \& Andersson(2021)]{GittinsRel2021} Gittins F., Andersson N., 2021, MNRAS, 507, 116
\bibitem[Gittins, Andersson \& Jones(2021)]{Gittins2021} Gittins F., Andersson N., Jones D. I., 2021, MNRAS, 500, 5570
\bibitem[Haskell, Jones \& Andersson(2006)]{Haskell2006} Haskell B., Jones D. I., Andersson N., 2006, MNRAS, 373, 1423
\bibitem[Haskell \& Melatos(2015)]{HaskellMelatos2015} Haskell B., Melatos A., 2015, Int. J. Mod. Phys. D, 24, 1530008
\bibitem[Horowitz \& Kadau(2009)]{HorowitzKadau2009} Horowitz C. J., Kadau K., 2009, Phys. Rev. Lett., 102, 191102
\bibitem[Johnson-McDaniel \& Owen(2013)]{JohnsonMcDaniel2013} Johnson-McDaniel N. K., Owen B. J., 2013, Phys. Rev. D, 88, 044004
\bibitem[LVK Collaboration(2026)]{LVKO4ab2026} LIGO Scientific Collaboration, Virgo Collaboration, KAGRA Collaboration, 2026, Narrowband searches for continuous gravitational waves from known pulsars in the first two parts of O4, preprint
\bibitem[Ogata \& Ichimaru(1990)]{OgataIchimaru1990} Ogata S., Ichimaru S., 1990, Phys. Rev. A, 42, 4867
\bibitem[Pearson et al.(2012)]{Pearson2012} Pearson J. M., Chamel N., Goriely S., Ducoin C., 2012, Phys. Rev. C, 85, 065803
\bibitem[Prix(2004)]{Prix2004} Prix R., 2004, Phys. Rev. D, 69, 043001
\bibitem[Seveso et al.(2016)]{Seveso2016} Seveso S., Pizzochero P. M., Grill F., Haskell B., 2016, MNRAS, 455, 3952
\bibitem[Strohmayer et al.(1991)]{Strohmayer1991} Strohmayer T., Ogata S., Iyetomi H., Ichimaru S., Van Horn H. M., 1991, ApJ, 375, 679
\bibitem[Ushomirsky, Cutler \& Bildsten(2000)]{Ushomirsky2000} Ushomirsky G., Cutler C., Bildsten L., 2000, MNRAS, 319, 902
\end{thebibliography}
\end{document}